\documentclass[sigconf]{acmart}
\AtBeginDocument{%
  }

\usepackage{xspace}
\usepackage{svg}
\usepackage{subcaption}

\newcommand{\eg}{e.g.,\xspace}
\newcommand{\etal}{et al.\xspace}
\newcommand{\ie}{i.e.,\xspace}

\newcommand{\sect}{Section\xspace}

\newcommand{\fig}{Figure\xspace}

\copyrightyear{2026}
\acmYear{2026}
\setcopyright{cc}
\setcctype{by}
\acmConference[CHI '26]{Proceedings of the 2026 CHI Conference on Human Factors in Computing Systems}{April 13--17, 2026}{Barcelona, Spain}
\acmBooktitle{Proceedings of the 2026 CHI Conference on Human Factors in Computing Systems (CHI '26), April 13--17, 2026, Barcelona, Spain}
\acmDOI{10.1145/3772318.3791228}
\acmISBN{979-8-4007-2278-3/2026/04}

\begin{document}

\title[]{A Collaborative Crowdsourcing Method for Designing External Interfaces for Autonomous Vehicles}


\author{Ronald Cumbal}
\email{ronald.cumbal@it.uu.se}
\affiliation{%
  \institution{Uppsala University}
  \department{Department of Information Technology}
  \city{Uppsala}
  \country{Sweden}
}
\author{Marcus G{\"o}ransson}
\email{marcus97goransson@gmail.com}
\affiliation{%
  \institution{Uppsala University}
  \department{Department of Information Technology}
  \city{Uppsala}
  \country{Sweden}
}
\author{Alexandros Rouchitsas}
\email{alexandros.rouchitsas@it.uu.se}
\affiliation{%
  \institution{Uppsala University}
  \department{Department of Information Technology}
  \city{Uppsala}
  \country{Sweden}
}
\author{Didem G{\"u}rd{\"u}r Broo}
\email{didem.gurdur.broo@it.uu.se}
\affiliation{%
  \institution{Uppsala University}
  \department{Department of Information Technology}
  \city{Uppsala}
  \country{Sweden}
}
\author{Ginevra Castellano}
\email{ginevra.castellano@it.uu.se}
\affiliation{%
  \institution{Uppsala University}
  \department{Department of Information Technology}
  \city{Uppsala}
  \country{Sweden}
}

\renewcommand{\shortauthors}{Cumbal et al.}

\begin{abstract}
Participatory design effectively engages stakeholders in technology development but is often constrained by small, resource-intensive activities. This study explores a scalable complementary method, enabling broad pattern identification in the design for interfaces in autonomous vehicles. We implemented a human-centered, iterative process that combined crowd creativity, structured participatory principles, and expert feedback. Across iterations, participant concepts evolved from simple cues to multimodal systems. Novel suggestions ranged from personalized features, like tracking lights, to inclusive elements like haptic feedback, progressively refining designs toward greater contextual awareness. To assess outcomes, we compared representative designs: a popular-design, reflecting the most frequently proposed ideas, and an innovative-design, merging participant innovations with expert input. Both were evaluated against a benchmark through video-based simulations. Results show that the popular-design outperformed the alternatives on both interpretability and user experience, with expert-validated innovations performing second best. These findings highlight the potential of scalable participatory methods for shaping emerging technologies. 
\end{abstract}




\maketitle

\section{Introduction}
\label{sec:Introduction}
Participatory methods have long been recognized as effective for involving stakeholders in the development of new technologies \cite{Bodker2022Participatory, simonsen2013routledge}. Yet, these methods are often resource-intensive and usually limited to small groups. In contrast, crowdworking platforms provide a scalable alternative, enabling distributed problem-solving through the web \cite{nakatsu2014taxonomy, morschheuser2017gamified}. Integrating participatory approaches into such platforms offers the potential to combine the benefits of user involvement with the breadth of large-scale participation.

Studies have shown that crowdsourcing can positively influence product design \cite{JIAO2021101496, DAHLANDER2014812, poetz2012value}, and recent work demonstrates that it can also generate valuable insights for user-centered technology design \cite{cumbal2025crowdsourcing}. 
However, while traditional crowdsourcing tasks typically generate pragmatic and functional outcomes, they often restrict participants' capacity for creative exploration. To overcome this limitation, we enhance this process though a collaborative and iterative design process that fosters imagination. By encouraging participants to build upon one another's contributions, we aim to stimulate collective creativity, enabling crowdworkers to envision diverse contexts and reflect on how their ideas might shape emerging technologies. In this way, human-centered design moves beyond documenting practical needs to serve as a speculative tool that empowers participants to imagine new technological futures. 

We apply this method to the design of external Human–Machine Interfaces (eHMIs) for Autonomous Vehicles (AVs).  Pedestrians depend on vehicle cues to feel safe, seeking acknowledgment similar to that from human drivers \cite{schieben2019designing}, and unclear signals create discomfort and uncertainty \cite{risto2017human, yang2017driver}. Research on eHMI design has therefore explored multiple modalities and message types, including support for diverse and accessibility-focused users \cite{dey2020taming, man2025pedestrians}. Although this work has expanded the design space significantly, consensus on effective strategies remains limited, and real-world variability continues to pose challenges \cite{dey2020taming}. This environment is well suited for large-scale participatory ideation, allowing participants to build on one another's insights, explore new interface concepts, and collectively assess communication strategies for clear and trustworthy AV–pedestrian interaction. This context motivates two research questions:

\begin{itemize}
\item \textbf{RQ1}: \textit{How does a collaborative, iterative crowdsourcing approach shape participants’ collective vision of eHMI design?}
\item \textbf{RQ2:} \textit{How do user-generated eHMI designs compare to prior ones in interpretability and user experience?}
\end{itemize}

Our contributions are threefold. \textbf{First}, we introduce a collaborative crowdsourcing method for engaging everyday users in technology design, demonstrated through eHMIs for AVs, where participants generated creative ideas (e.g., robot drivers, predictive trajectories) shaped by expert feedback. \textbf{Second}, we offer insights into user expectations, showing a consistent preference for familiar, standardized signals. \textbf{Finally}, we validate our approach by comparing crowdsourced and benchmark designs ---limited to visual elements--- finding that the most frequently suggested concepts outperform alternatives in both interpretability and user experience, with expert-validated innovations performing second best.

\section{Positioning Our Approach in Relation to Participatory Design (PD)}
\label{sec:PositioningOurApproachInPD}
Our motivation for using a participatory approach stems from the belief that user involvement is essential in shaping future technologies \cite{Bodker2022Participatory}. Our approach, however, does not seek to replicate participatory design in full. Instead, it draws selectively on its core commitments, including \textit{democratization}, \textit{mutual learning}, and the recognition of \textit{people as skillful contributors} \cite{bannon1995human, Bodker2022Participatory}. While these commitments inform our work, we also acknowledge that our method operates at different depths of engagement and power dynamics.

Our application of eHMI design is a particularly suitable case because the field is still emerging, lacks standardized practices, and may strongly benefit from user-centered approaches to address these gaps \cite{dey2020taming, man2025pedestrians, Mandujano2024}. In this context, the principles we borrow from PD help us engage diverse populations to better understand user needs \cite{CARROLL2007243}, although we recognize that our process cannot build the forms of community that characterize traditional PD.

By inviting \textit{participants to explore alternative imaginaries}, we leverage collective creativity to surface ideas that might not emerge in small, co-located groups. 
Incorporating \textit{large numbers of stakeholders} enables rapid identification of stable preference patterns and can reduce uncertainty. 
\textit{Mutual learning} is supported both among participants, who can build on each other's contributions, and between participants and researchers, who exchange knowledge about emerging technologies through expert feedback. At the same time, we recognize that the power structures and democratization central to participatory design remain only partially addressed. Crowdsourcing platforms introduce well-known sampling biases \cite{brewer2016would, Difallah2018, Kotaro2018, uzor2021investigating}, limiting representation and constraining empowerment. Mutual learning can occur, but its depth is necessarily restricted because online participation cannot sustain the continuous, situated interaction that characterizes in-person PD.

Furthermore, unlike conventional surveys, our process enables bidirectional information flow and supports creativity through iterative visualization. Traditional survey definitions typically position participants only as sources of information, rather than as recipients of knowledge \cite{fink2024conduct}. \textit{Our approach occupies a complementary space between traditional PD methods and large-scale data collection.}  It does not aim to reproduce the embodied, relational knowledge of face-to-face PD sessions, nor the one-way information extraction characteristic of surveys (see Table \ref{app:MethodologicalComparison} for a detailed comparison).

\section{Related Work}
\label{sec:RelatedWork}

\subsection{Participatory Design of eHMIs}
In the context of eHMIs, large-scale surveys have yielded important insights into public perceptions and expectations \cite{ammar2024building}, though with the previous mentioned limitations. Co-design sessions have been used to enable deeper involvement \cite{verma2019pedestrians, lakhdhir2023wearing, chauhan2024transforming}, but face practical constraints, including small sample sizes and high resource demands. Notable PD efforts include CityMobil2, which combined interviews, on-site questionnaires, and focus groups to derive actionable recommendations for automated minibuses \cite{schieben2019designing}. Other studies foreground underrepresented groups, such as people with mobility restrictions \cite{Asha_2020, asha2021co} or visual impairments \cite{colley2020towards}, highlighting persistent barriers to accessible AV communication. 
Of particular relevance to our approach, Mahadevan \etal \cite{mahadevan2018communicating} used the PICTIVE method to enable participants to generate 34 unique eHMI concepts. Lakhdhir et al. \cite{lakhdhir2023wearing} later extended this approach to \textit{pedestrian-wearables}, illustrating how personal accessories could serve as eHMIs. 
These studies illustrate the benefits of involving pedestrians during the ideation phase, yet their smaller sample sizes restrict the robustness and generalizability of their findings. This challenge opens opportunities for a complementary scalable approach. 

\begin{figure*}[!t]
  \centering
  \includegraphics[width=\linewidth]{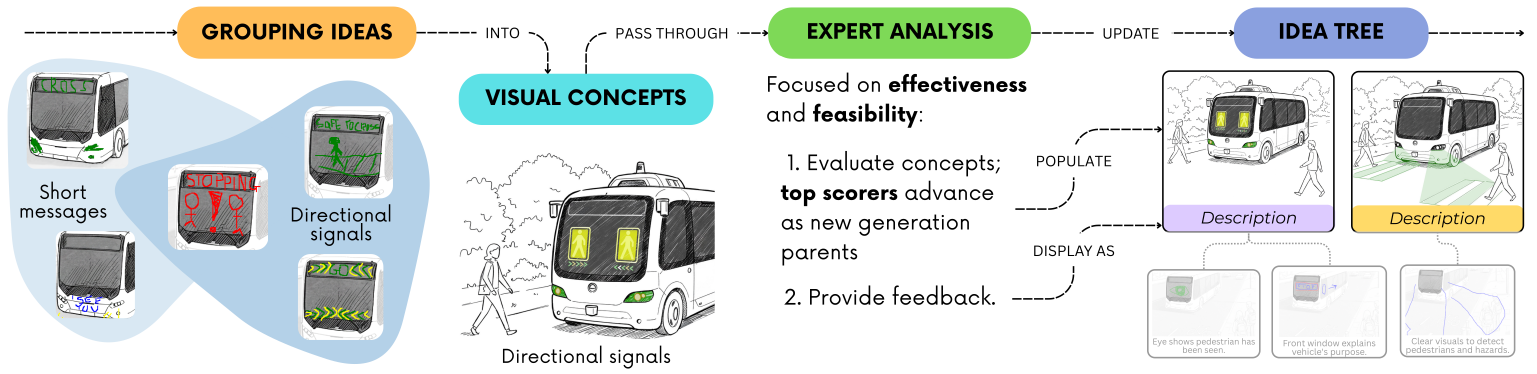}
  \caption{Illustration of the \textit{Expert Analysis and Feedback} process. Participant-submitted sketches are grouped into similar clusters (\textit{concepts}). An independent expert rates each concept's effectiveness and feasibility, providing justifications. These illustrated concepts, with descriptions and feedback, form the next stage of the visualization tree, guiding the next iteration.}
  \Description{The figure shows a process chart beginning with Grouping Ideas on the left, illustrated with examples of participants' sketches. This leads to Creating Visual Concepts, where standardized versions of the sketches are shown. Next is the Expert Analysis step, accompanied by the text ``Focused on effectiveness and feasibility.'' The process concludes with the Idea Tree, displaying a small tree that organizes the concepts visually. }
  \label{fig:figure1}
\end{figure*}

\subsection{Crowd-Based Design}
Crowdsourcing harnesses distributed online communities to solve general problems or tasks \cite{nakatsu2014taxonomy, morschheuser2017gamified}. In design contexts, crowd feedback has been shown to enhance creative output \cite{morris2013affect, sun2015collaborative}. For instance, Xu \etal \cite{xu2015classroom} demonstrated that non-expert feedback helped design students improve their work at superficial and conceptual levels. However, expert validation remained essential for ensuring design feasibility. Similarly, Park \etal \cite{park2013crowd} found that combining crowd and expert contributions in competitive, collaborative settings led to higher-quality results and increased stakeholder satisfaction. 
Despite its promise, crowd-based design has seldom been applied to more technical domains, where participant expertise is expected to be limited. Nonetheless, we argue that combining crowdsourcing with structured guidance and expert feedback can bridge this gap, leveraging the crowd's imagination while ensuring technical feasibility.

\subsection{Communicating Awareness with eHMIs}
A central function of eHMIs is to support pedestrians' decision-making by making vehicle detection and intent understandable and trustworthy. Prior work shows that pedestrians consistently prioritize confirmation that the vehicle has detected them \cite{schieben2019designing, merat2018externally, rouchitsas2019external}, followed by cues that communicate the vehicle's intent, such as yielding or proceeding \cite{mahadevan2018communicating, lakhdhir2023wearing}. 
To address these needs, researchers have explored diverse communication strategies, including 360° displays \cite{verstegen2021commdisk}, light strips \cite{nissan2015, schlackl2020ehmi}, icons \cite{eisele2025effects}, anthropomorphic cues like eyes or facial expressions \cite{jaguar2018, chang2017eyes, verstegen2021commdisk, rouchitsas2023smiles}, text-based messages \cite{epke2021see}, and multimodal combinations \cite{mahadevan2018communicating}. These approaches vary in interpretability and scalability: for instance, displays designed to address multiple pedestrians often struggle in dense environments, where individual attention is difficult to manage \cite{dey2020taming}. 
This body of work highlights the ongoing challenge of balancing clarity, scalability, and contextual adaptability, motivating the continued exploration of user-centered creative eHMI concepts.

\section{Method: Collaborative Crowdsourcing for eHMI Design in Autonomous Vehicles}
Our method integrates \textit{participatory design principles} with scalable \textit{crowdsourcing techniques} to combine meaningful stakeholder engagement with creative exploration, as explained in Section \ref{sec:PositioningOurApproachInPD}. 
The creative objective is structured around the following components:

\begin{itemize}
\item Iterative Idea-building: Adapting Yu et al.'s approach \cite{yu2011cooks}, participants expand on and combine prior submissions, fostering more original and higher-quality design outcomes. To further support this process, we incorporate expert feedback to guide and refine participant contributions. 
\item Structured Visualization: Following Sun et al.'s tree visualization strategy \cite{sun2015collaborative}, we use a tree-based visualization to organize the evolution of ideas, maintaining clarity across iterations. We integrate expert feedback to bridge the gap between creative exploration and practical implementation. 

\end{itemize}

\begin{figure*}[!t]
  \centering
  \includegraphics[width=\linewidth]{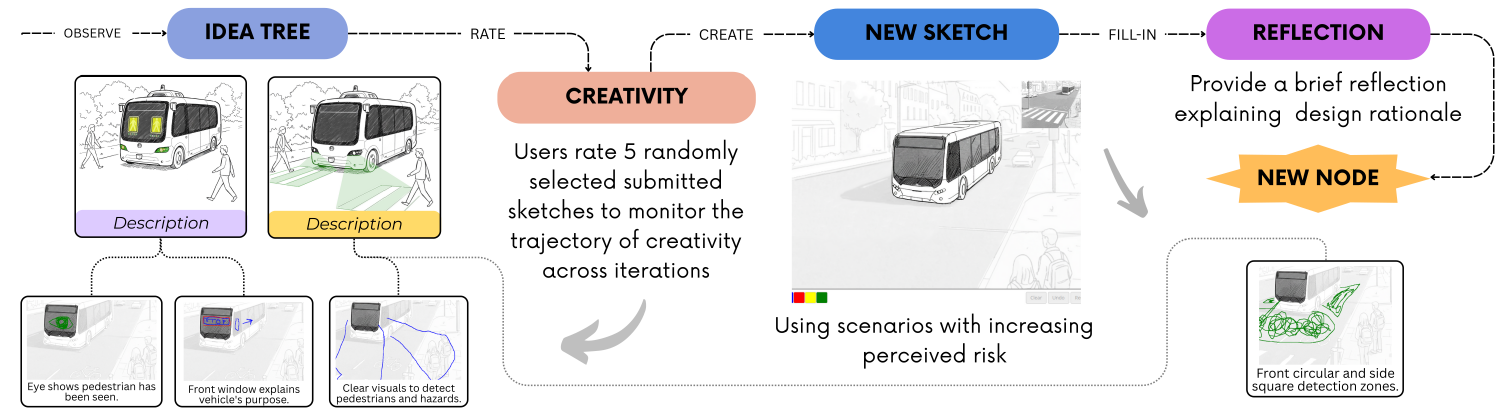}
  \caption{Illustration of the \textit{Crowdsourced Design Collaboration} process. Participants iteratively refine ideas inspired by earlier contributions. Using a tree-based visualization, top-level concepts branch into subsequent sketches. Participants review and rate prior concepts and then submit their contribution. Each submission includes a sketch and a brief explanation.}
  \Description{The figure presents a process chart starting with the Idea Tree, which displays a small tree organizing the concepts visually. This is followed by Creativity Rating, where users evaluate five randomly selected submitted sketches. Next is the New Sketch step, showing the interface for drawing new ideas. The process then moves to Reflection, which includes the instruction ``Provide a brief reflection explaining the design rationale.'' Finally, a New Node is created, illustrated with a sample submitted sketch and its accompanying description.}
  \label{fig:figure2}
\end{figure*}


Furthermore, our method alternates between two stages: (1) crowdsourced design collaboration (\sect~\ref{sec:CrowdosurcedDesignCollaboration}) and (2) expert analysis with feedback (\sect~\ref{sec:ExpertAnalysisFeedback}). While the process could have begun directly with participant collaboration, we first evaluated eHMI sketches from prior research to establish a baseline of widely studied design elements. These included concise \textit{text messages}, dynamic \textit{light displays}, directional light \textit{projections}, and \textit{sound} cues, which were organized as the initial concepts at the top of the visualization tree (see \fig~\ref{fig:figure10} in Appendix \ref{app:FinalConceptVisualizationTree}). This was also intended to provide grounding in eHMI design without creating strong priming.\footnote{In pilot studies, explicitly showing previous eHMI designs caused participants to replicate existing ideas without much creative input.}

\subsection{\textbf{Expert Analysis and Feedback}}
\label{sec:ExpertAnalysisFeedback}
Participant-generated sketches were clustered into conceptually similar groups, with each cluster defined as a \textbf{concept} and standardized for consistent graphical representation. An independent expert\footnote{Concepts were evaluated by one of the authors, an independent expert in Human Factors of Automated Driving, who was blind to the study methodology and goals.} assessed each concept on two criteria: \textit{effectiveness} (clarity of pedestrian awareness communication) and \textit{feasibility} (practicality of implementation). These evaluations were recorded using a 7-point Likert scale, each accompanied by short justifications. Summarized versions of the feedback were then embedded into the visualization tree as parent nodes for subsequent iterations (as illustrated in \fig~\ref{fig:figure1}).  

\subsection{\textbf{Crowdsourced Design Collaboration}}
\label{sec:CrowdosurcedDesignCollaboration}

Building on the expert-reviewed concepts, participants engaged in collaborative cycles of refinement, combination, and reimagination. This process follows the human-based genetic algorithm framework proposed by Yu \etal \cite{yu2011cooks}, visualized through a tree structure adapted from Sun \etal \cite{sun2015collaborative}. In this structure, concepts---paired with expert feedback---form the top layers, while participants' sketches branched below.

The website used to collect participant contributions was developed using \href{https://www.jspsych.org/v7/}{jsPsych} and hosted on \href{https://www.cognition.run/}{Cognition}. Participants first approved consent and completed a short demographic questionnaire (age, gender, primary transportation mode, and AV familiarity), followed by a brief introduction that highlighted the study's goals and the importance of AV–pedestrian communication.

To encourage participants to reflect on different aspects of AV– pedestrian communication, we employed two scenarios that differed in perceived risk, defined from the pedestrian's point of view in line with prior eHMI research \cite{de2023predicting, ha2020effects}. 

\begin{itemize}
    \item \textbf{Low-Risk Scenario}: A pedestrian crosses at a signalized zebra crossing as an autonomous vehicle approaches. Because the AV is expected to follow the traffic light \cite{schieben2019designing}, the situation feels safe and predictable.
    \item \textbf{High-Risk Scenario}: A pedestrian crosses in a crowded area without clear traffic signals. The AV is closer, creating more uncertainty.
\end{itemize}

Participants were initially introduced to the low-risk traffic scenario and asked, ``\textit{How would you like the AV to communicate with you?}" This initial prompt encouraged them to engage with the task in a concrete, real-world context. They were then shown the visualization tree, which presented the most common elements used in previous eHMIs in the top layer. After this, participants rated five randomly selected designs on creativity using a 7-point Likert scale, enabling us to monitor changes in creativity across iterations.

Next, participants were introduced to the high-risk scenario alongside seven randomized reflective prompts (\eg \textit{``What if two children step onto the street from opposite sides in heavy rain?''}) to encourage the consideration of real-world complexities. After revisiting the visualization tree one more time, they created their own design using the embedded drawing interface and provided a short written explanation of their rationale. The reflections offered contextual insight into how participants' reasoning shaped their design contributions (RQ1). Participants also recorded whether the ideas were built upon existing concepts or were completely new. 

We applied a \textit{stopping criterion} to determine when to conclude data collection. Each submission was reviewed by the main author of this work to assign a novelty score (5 = new modality, 3 = substantial refinement, 1 = minor tweak) and check for the introduction of new signaling modalities. Data collection stopped once the average novelty score of the last five participants fell below 2 and no new modality appeared in three consecutive submissions.

\begin{figure*}[t]
  \centering
  \includegraphics[width=0.99\linewidth]{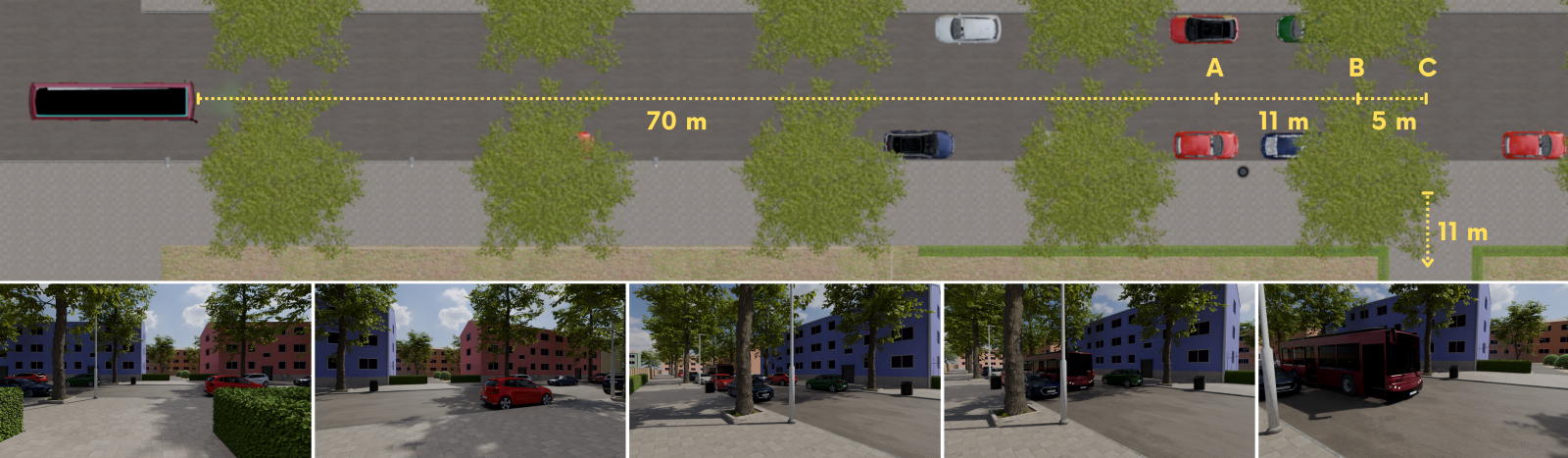}
    \caption{Configuration of the traffic scenario used for evaluating eHMI designs. Key points where specific cues were activated are highlighted: (Point A) communication of pedestrian detection, (Point B) signaling yielding intention, and (Point C) additional crossing facilitation. Sample screenshots from the simulation of an autonomous bus without eHMI are shown at the bottom.}
    \Description{The image provides a top-down view of a pedestrian and a bus as they approach a crossing intersection. A geometric overlay illustrates the measurements of the scenario, depicting the spatial constraints and positioning of both the pedestrian and the bus.}
  \label{fig:figure3}
\end{figure*}

\section{Method: Final Concept Evaluation}
\label{sec:MethodFinalConceptEvaluation}
We evaluate the effectiveness of concepts generated through our collaborative approach using established behavioral response experiments, in which participants view video stimuli of simulated AVs featuring different eHMI designs. While video-based studies lack full interactive immersion and real-world risk \cite{man2025pedestrians}, they remain a widely used, safe, and cost-effective alternative to field tests \cite{Brade2017, dey2020color}. Importantly, we focus on testing the visual components only since they are the most practical to evaluate in an online setting. Controlling for other modalities, such as volume or room acoustics for audio, is much more complex through crowdsourcing platforms. However, eHMIs are inherently multimodal, and future work should extend this evaluation accordingly.

\subsection{\textbf{Experiment Conditions}}
\label{sec:ExperimentConditions} 
To assess the effectiveness of our collaborative crowdsourcing approach, we compare these user-generated eHMI designs with baseline concepts from prior research (RQ2). This comparison enables us to evaluate both the extent to which everyday (crowdsourced) participants can meaningfully contribute to the design of emerging technologies and whether their designs meet user expectations and perform comparably to established solutions.

\subsubsection{\textbf{Baseline eHMI}}
We selected a baseline design that represents widely adopted eHMI approaches, ensuring comparison with proven designs rather than arbitrary ones. First we selected the $360^\circ$ display with ``eyes'' proposed by Verstegen \etal \cite{verstegen2021commdisk}. This awareness-focused design has been recognized in multiple studies as scalable and highly visible, including in interactions with vulnerable road users \cite{yu2024understanding, Al2024light, wang2022pedestrian, kang2025you, BERGE2023104043, alhawiti2024effectiveness}. 

The second element selected was a light band, which Dey \etal \cite{dey2024multi} identified as the ``\textit{most widely used/proposed visual eHMI due to its relative simplicity, ease of implementation, and abstract execution}'', with several studies applying this approach \cite{dey2020taming, ACKERMANN2019272, de2019external, Dey2018Interface, Faas2020Yielding, habibovic2018communicating, hamm2018ideas, Hensch2020How, petzoldt2018potential}. 
This combination represents current best practices: the light band is the most prevalent concept in the literature, while the ``eyes'' convey display directionality.

\subsubsection{\textbf{Crowdsourced eHMIs}}
From the wide range of ideas proposed by participants, we selected two representative designs. First, a \texttt{popular-design} that captures the solutions most frequently proposed during the iterative process, directly reflecting participants' shared expectations and intuitions. Second, an \texttt{innovative-design} that integrates creative user concepts with expert evaluation, balancing novelty and feasibility. This design is derived by averaging expert ratings on effectiveness and practicality to identify the highest-scoring concepts.

\subsection{Measurements}
We evaluated the designs using two widely applied criteria in eHMI research \cite{man2025pedestrians}. First, \textbf{interpretability} was assessed through participants' \textit{response times}. Response time is commonly used as a proxy for comprehension speed; shorter times indicate that participants understand the eHMI signal more quickly. In practice, participants press a key when they feel safe to cross in front of an approaching vehicle, as in prior head-mounted display studies \cite{de2019external} and video-based experiments \cite{noomwongs2025evaluating, ONKHAR2022194, eisma2019external}. To complement this objective measure, we collected participants' subjective impressions. These were assessed using the short version of the \textit{\textbf{User Experience} Questionnaire} (UEQ-S) \cite{hinderks2017design}, administered after each video stimulus. The UEQ-S has been effectively applied in both video-based \cite{haimerl2022evaluation} and VR-based eHMI studies \cite{yue2025enhancing}. We note that willingness to cross reflects both comprehension speed and perceived safety or comfort, and the UEQ-S therefore provides valuable complementary insight into user experience beyond interpretability alone. 

\subsection{Hypotheses}
Since user-centered design methods emphasize intuitive, relatable interactions, we expect crowdsourced eHMIs to outperform the baseline. Specifically, we hypothesize that (\textbf{H1}) \textit{crowdsourced eHMI designs will score higher than the baseline in interpretability (\textbf{H1.a}) and user experience (\textbf{H1.b})}. 
Moreover, the integration of expert feedback ensures that participant-generated designs remain aligned with principles established in the eHMI research community. Accordingly, we hypothesize that (\textbf{H2}) \textit{the \texttt{innovative-design} will be the most easily interpreted in conveying an AV's pedestrian awareness, outperforming all other eHMI designs}.

\subsection{Simulation Environment}
\label{sec:SimulationEnvironment}
We created videos in \href{https://www.blender.org/}{Blender}, using both self-made and publicly available assets (e.g., vehicles, vegetation). The simulated environment depicted a high-risk scenario characterized by the absence of traffic signals and occluded visibility in a crowded setting (\fig \ref{fig:figure3} shows a top view of this configuration). A total of four 15-second videos were produced: one with the AV without any eHMI (used as a practice run) and three videos corresponding to the experimental conditions described in \sect \ref{sec:ExperimentConditions}. In each video, the scenario remained consistent: a bus begins moving about 85 meters from the pedestrian, traveling at 36 km/h\footnote{The average straight line speed for buses in the European Union is 14.5 km/h \cite{poelman2020many}. }, and decelerating at 20 meters until stopping in front of the pedestrian. The camera followed the pedestrian's perspective, walking at 1.27 m/s. Before approaching the crossing, the pedestrian slightly turned left and right to simulate scanning traffic, until it stood close to the sidewalk border. 

\subsection{Experimental Procedure}

The experiment was implemented using the jsPsych library and hosted on Cognition\footnote{Code made publicly available through a \href{https://github.com/ronaldcumbal/chi2026_collaborative_crowdsourcing}{GitHub repository.}}. Participation was restricted to devices with a minimum screen resolution of 1280×720 pixels. After providing consent, participants reported demographic information including age, gender, primary transportation mode, and acceptance of AVs.

For the main task, participants were instructed to press the \texttt{Enter} key at the moment they felt it was safe to cross the street while watching each video stimulus. To familiarize them with the procedure, two practice trials were presented. First, a low-risk scenario with a different vehicle, to make sure participants were pressing the correct key. 
The main session then began with the video of a vehicle without an eHMI, after which participants went through the randomized sequence of videos featuring the three eHMI experimental conditions. After each video, participants completed the \textit{User Experience Questionnaire} and described their interpretation of the design in a text box. At the end of the study, they could provide open-ended feedback in a similar text. No prior explanation of the eHMIs was given, ensuring spontaneous and intuitive responses, consistent with previous video-based studies \cite{eisma2019external, oudshoorn2021bio}.

\begin{figure*}[t]
  \centering
  \includegraphics[width=1\linewidth]{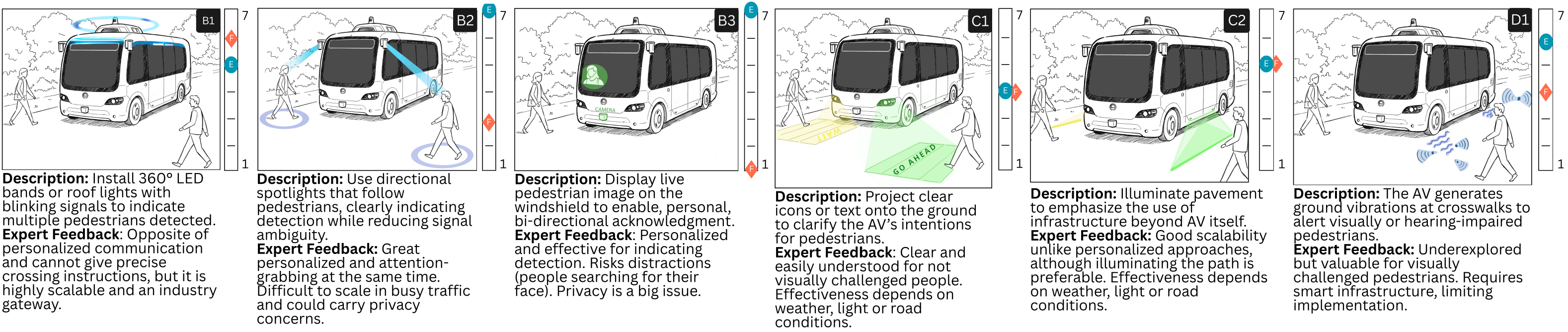}
  \caption{Concepts developed in the \textbf{first} collaborative iteration. All concepts have a brief descriptions and expert feedback, along with expert ratings for effectiveness ($\circ$) and feasibility ($\diamondsuit$), where 7 indicates the highest effectiveness and feasibility.}
  \Description{The figure illustrates the various concepts developed across iterations. Concept B1 features a blue 360-degree light band. Concept B2 displays beam lights directed at pedestrians, encircled in purple light. Concept B3 incorporates a green human face within the same frame as a pedestrian. Concept C1 shows green and yellow projections on the ground with text. Concept C2 presents green light projections along the side of the street. Concept D1 uses squiggly lines to represent tire vibrations from the bus.}
  \label{fig:figure4}
\end{figure*}

\section{Results: Collaborative Design with Expert-Guided Iterations}
\label{sec:Results_Collaboration_Feedback}

Participants were recruited through \href{https://www.prolific.com/}{\texttt{Prolific}} and had to meet eligibility criteria of European residence and English proficiency. This ensured a population with a similar context in terms of urban environments and mobility infrastructure. No prior design experience was required, as our goal was to capture everyday pedestrians' perspectives as end-users of eHMI technology. A total of 67 participants submitted designs sketches throughout four process iterations. The participants had an average age of 37.86 (SD: 11.82), with gender distribution of 1 non-binary, 31 male, and 35 female participants. Out of these, 42 used a private vehicle for transportation, 14 Public transportation and 11 Walking/Cycling. 
Regarding prior interaction with AVs, 49 said ``No,'' 13 responded ``Yes,'' and 5 were ``Not sure.''  The average response to using self-driving vehicles when they become available (\ie technology acceptance) was slightly more positive than "Not Sure" (M: 3.33 SD: 1.07). Participants were from 13 European countries, with the United Kingdom being the largest sample (38), then Poland (7), Italy (4), and Germany (4) accounting for the majority. 
The median study duration was 17 minutes, and participants were compensated at \pounds $12.91$ per hour.

\subsection{Thematic Analysis}
Participant submission were analyzed using reflexive thematic analysis \cite{braun2006using}\footnote{Code and themes made publicly available through a \href{https://github.com/ronaldcumbal/chi2026_collaborative_crowdsourcing}{GitHub repository.}}. The first author conducted open coding to identify recurring ideas. This analysis focused mainly on the reflective and descriptive responses provided for the high-risk questions associated with the sketch. The low-risk answers were used only to clarify context when needed. These codes were then grouped into themes reflecting both the design priorities expressed in the sketches and the motivations behind participants' reasoning. For design elements, eighteen (18) codes were generated and organized into five themes: \textit{lights}, \textit{displays/screens}, \textit{road projection}, \textit{audio}, \textit{novel ideas} and \textit{sensing}
\footnote{We do not report on \textit{sensing} as this work focuses on how awareness is communicated, rather than on the sensor capabilities required to achieve it.}.
For motivation codes, seven (7) themes were discovered: \textit{adaptiveness}, \textit{awareness}, \textit{familiar conventions}, \textit{accessibility}, \textit{intent transparency}, \textit{safety}, and \textit{visibility}. See Appendix \ref{app:ThematicAnalysis} for the thematic code distribution using \href{https://atlasti.com/}{Atlas.ti}.

\textbf{Note:} In the following results, we report the number of participants who mentioned each theme in each interaction round, counting multiple mentions by the same participant only once.

\begin{figure*}[t]
  \centering
  \includegraphics[width=1\linewidth]{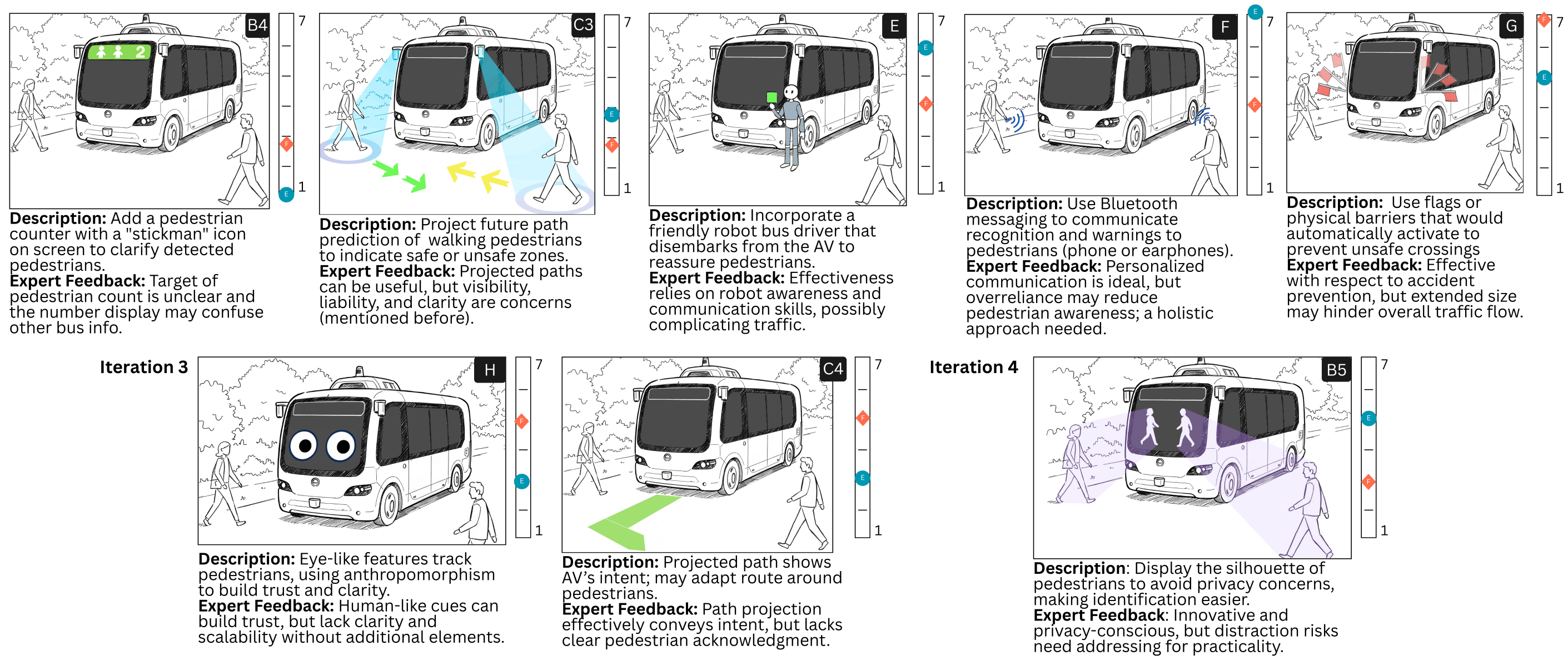}
  \caption{Concepts developed in the \textbf{second}, \textbf{third} and \textbf{fourth} iterations. All concepts have brief descriptions and expert feedback, along with expert ratings for effectiveness ($\circ$) and feasibility ($\diamondsuit$), where 7 indicates highest effectiveness and feasibility.}
  \Description{The figure depicts various concepts developed across iterations. Concept B4 features a pedestrian counter in green on the top front of the bus. Concept C3 shows a light beam directed toward participants, accompanied by yellow and green arrows indicating the predicted path. Concept E presents a robot in front of the bus holding a green card. Concept F includes sound icons above pedestrians to represent Bluetooth communication. Concept G displays red flags dropping from the sides of the front of the bus. Concept B5 shows icons shaped like the participants on the front of the bus. Concept C4 depicts a green arrow illustrating the future path of the bus, and Concept H features eyes on the front of the bus. }
  \label{fig:figure5}
\end{figure*}

\subsection{\textbf{Iteration 1}}
We first recruited 26 participants, using creativity ratings as the stopping criterion\footnote{Due to a bug, the last five submissions were not randomly selected for tracking creativity levels and were therefore excluded from analysis.}. However, the ratings showed no clear peaks or declines, making it difficult to determine when creativity was diminishing (see \fig \ref{fig:figure8}). Instead, we observed that most novel modality expansions, such as tactile signals, projections, and personalized displays, appeared between the 7th and 10th participants. After the 12th to 15th submissions, ideas increasingly shifted toward incremental refinements (e.g., color changes, added sounds) rather than introducing new concepts. Based on these patterns, we revised our stopping criterion to combine novelty scores with new modality annotations, as described in \sect \ref{sec:CrowdosurcedDesignCollaboration}.

Analysis of the submissions revealed several novel concepts (illustrated in \fig \ref{fig:figure4}) as well as four main submission categories. The first, and most common, was light-based signals (18/26 participants), already grouped under \textit{Concept B}. Participants suggested dynamic, color-coded lights to improve visibility, with some proposing multiple blinking lights to signal multiple pedestrians. Three distinct variations emerged: \textbf{Concept~B1}, omnidirectional $360^{\circ}$ LED bands or roof lights (P17, P21); \textbf{Concept B2}, directional spotlights tracking pedestrians (P14); and \textbf{Concept B3}, live pedestrian images displayed on the AV's external screen (P8) as a form of bi-directional acknowledgment. 
The next category focused on road and pavement projections (10/26 participants), where participants suggested projecting signals directly onto the street or illuminating surrounding pavement (grouped under \textit{Concept C}). Two new approaches were identified: \textbf{Concept C1}, projecting icons or text messages (P15); and \textbf{Concept~C2}, directional lighting on nearby infrastructure (P12, P13). 
The other category included sound or vibrating signals (11/26 participants), conforming to \textit{Concept D}, such as sirens and verbal alerts to attract attention and assist visually impaired pedestrians. One innovative proposal, \textbf{Concept D1}, used ground vibrations at crosswalks (P25) to provide tactile warnings, offering an inclusive solution in noisy or low-visibility environments. Finally, participants also mentioned front displays that could indicate the vehicle's intentions or provide other information (12/26 participants), a feature that was also described in \textit{Concept~A}. 

\subsection{\textbf{Iteration 2}}
This iteration included 19 participants and largely mirrored the patterns observed earlier. Lights remained the dominant communication method (15/19 participants).  
Other elements frequently proposed included sound, like recorded messages, beeps, or music (8/19 participants).  
Projecting visuals onto the ground or road were also mentioned (9/19 participants).  
Two novel concepts also emerged: the use of dynamic safe/unsafe zones around the vehicle using dual-color lights (P27, P44) and future path prediction to indicate safe or unsafe zones in advance (P28, P37), both defined under \textbf{Concept C3}. 
Displays were also suggested as a way to convey the AV's intentions (7/19 participants). \textbf{Concept B4} extended previous ideas by adding a pedestrian count and a ``stickman" icon (P35). Several unique proposals appeared: \textbf{Concept E}, a robot bus driver that exits the vehicle to reassure pedestrians (P29, P33); \textbf{Concept F}, Bluetooth messaging to deliver recognition and warnings (P42); and \textbf{Concept G}, physical interventions such as flags or barriers that automatically deploy to prevent unsafe crossings (P31, P39).

\subsection{\textbf{Iteration 3}}
This iteration was completed with 15 participants; however, the data from one participant was excluded due to responses that reflected an extremely negative---and at times offensive---stance toward the implementation of autonomous vehicles.

Light-based signals remained the most common proposal (13/14 participants). 
One participant (P46) suggested adapting light intensity based on pedestrian distraction, which we categorized as a general enhancement applicable across signal types. Audio cues also featured prominently (6/14 participants), with particular emphasis on voice clarity. 
Projections continued to appear (3/14 participants), and one participant proposed displaying the AV's future path to signal intent (P53). Although not directly tied to pedestrian awareness, this was introduced as \textbf{Concept C4}. 
Other recurring ideas involved icons, symbols, and text, extended to side-mounted displays (6/14 participants). \textbf{Concept H} was suggested as an eye-like animation to convey the AV's awareness of pedestrians (P56).

\subsection{\textbf{Iteration 4}}
The final iteration, after applying stopping criteria, involved seven participants and reinforced earlier trends. Lights remained the primary communication tool (6/7 participants), with sound or voice cues also suggested (4/7 participants), particularity to enhance clarity, though one participant (P66) noted concerns about noise pollution despite their usefulness. Road/environmental projections (3/7 participants) and pedestrian-type adaptations also persisted. Participants also proposed more targeted light-based approaches: spotlights or lasers to mark individuals (P62), like \textit{Concept B2}, and light fields to indicate safety zones (P65), similar to \textit{Concepts C1/C2}. \textbf{Concept B5} corresponds to a particularly interesting suggestion (“wall of light”), which projects pedestrian silhouettes as a privacy-friendly visualization to reassure pedestrians detection (P64).

\begin{figure*}[t]
  \centering
  \includegraphics[width=0.9\linewidth]{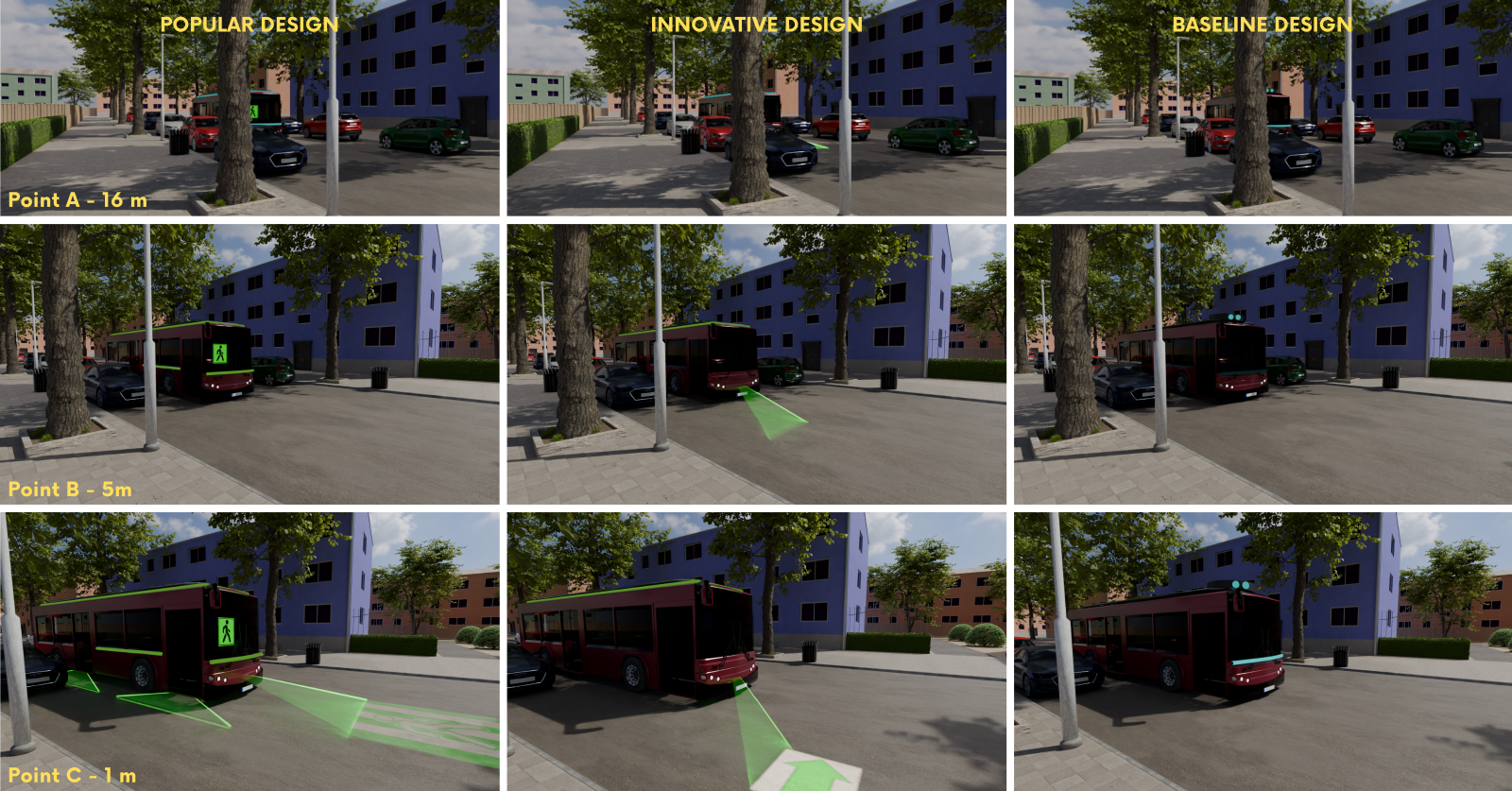}
    \caption{Screenshots of the three eHMI designs created in simulation captured over the three main points of AV-Pedestrian communication: (Point A) Pedestrian Awareness, (Point B) Yielding Intention, and (Point C) Crossing Facilitation. }
    \Description{The figure presents a grid of screenshots from the simulation environment. The left column shows the popular design, featuring a green pedestrian icon and green lights. The middle column displays the innovative design, with a green projected light and an arrow. The right column shows the baseline design, including a 360-degree disk at the top with eyes and a cyan light on the front.}
  \label{fig:figure6}
\end{figure*}

\section{Results: Final Crowdsourced eHMI Designs}
\label{ResultsFinalDesigns}
As outlined in \sect \ref{sec:ExperimentConditions}, we derived two final eHMI designs through our collaborative crowdsourcing approach. These designs are central to addressing our research questions, as they provide a concrete basis for empirical evaluation. The \texttt{popular-design} reflects the most frequently proposed ideas, capturing the collective vision that emerged through iterative refinement (RQ1). The \texttt{innovative-design} combines participant-generated concepts with expert feedback, illustrating how structured collaboration can move beyond common solutions toward novel yet practical approaches. For comparison, we included a \textit{baseline} design representing a state-of-the-art benchmark. Alongside our two crowdsourced concepts, this allowed us to examine how the three designs perform with respect to interpretability and user experience (RQ2). The designs are shown within the simulation environment in \fig~\ref{fig:figure6}.

\subsection{\textbf{Popular-design}}
As described in the results of the creative process (\sect~\ref{sec:Results_Collaboration_Feedback}), light signals consistently emerged as the dominant element through all collaborative iteration. The most frequently proposed idea was a \textit{360° light band} that changes color and pattern (steady green for ``safe to cross,''). 
Participants also frequently suggested \textit{ground projections} with high-contrast symbols and \textit{auditory cues}, ranging from calm tones in low-risk situations to louder beeps in high-risk contexts. A further recurring feature was \textit{pedestrian count and tracking}, which displayed the number of detected pedestrians. 

Since our study uses videos of simulated environments, we are limited to visual elements only, which excludes the use of auditory cues. 
As a result, the final design incorporates the \textit{360° light band} (green for ``safe to cross,") and \textit{icon display} to signal pedestrian awareness, along with the \textit{ground projections} of a ``zebra'' crosswalk to indicate a safe path. 

\subsection{\textbf{Innovative-design}} 
\label{sec:ResultsFinalDesigns-InnovativeDesign}
To identify a design that balanced participant creativity with expert evaluation, we averaged effectiveness and feasibility ratings across concepts and selected the highest-scoring novel ideas proposed by participants. This process identified several concepts, including the \textit{360° LED band} (Concept B1), a pedestrians' \textit{future path projection} (Concept C3), enabled \textit{Bluetooth communication} (Concept F) and the deployment of \textit{physical barriers} (Concept G). Because our study focused on visual communication, Bluetooth communication was excluded. The final design combined a green \textit{360° LED band} to indicate yielding with a \textit{future path projection} to convey pedestrian awareness, complemented by \textit{directional spotlights} (Concept B2), also rated highly , to further strengthen communication.

\begin{figure*}[t]
    \centering
    \includegraphics[width=0.92\linewidth]{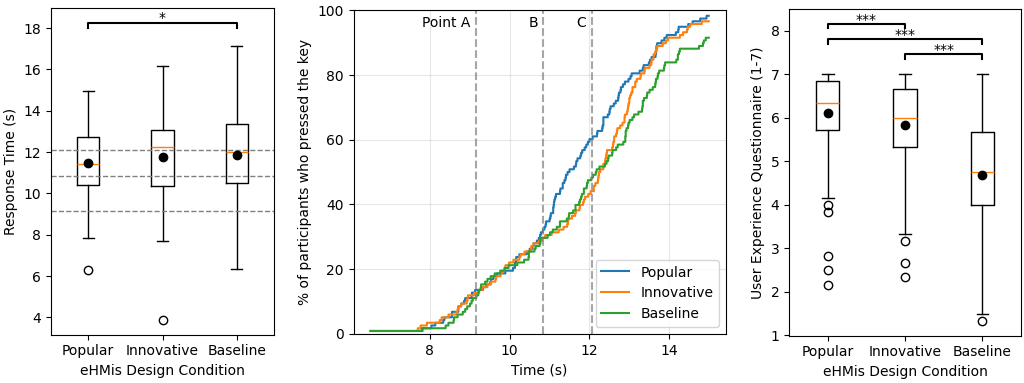}
    \caption{Results for response times and user experience. Significance levels indicated as ***= p<0.001 and *= p<0.05. (Note on the participant key press plot: a higher percentage of participants pressing the key earlier indicates higher interpretability.}
    \Description{The left plot presents a boxplot of response times, with all media showing similar heights, and a horizontal line indicating a significant difference between the popular and baseline designs. The middle plot illustrates the time evolution of the percentage of participants who pressed the key during the experiment. The right plot displays a boxplot of the user experience questionnaire scores, with declining medians and horizontal lines highlighting significant differences among all three conditions.}
    \label{fig:figure7}
\end{figure*}

\subsection{\textbf{Baseline Design}} 
As outlined in \sect \ref{sec:ExperimentConditions}, the \texttt{baseline} eHMI combines a \textit{360° display} with eye-inspired circles that move toward detected pedestrians, using multiple ``eyes'' when several pedestrians are present \cite{verstegen2021commdisk}. These circles blink to signal acknowledgment, while a small icon beneath them indicates the number of pedestrians. In addition, a cyan \textit{light band} wraps around the front of the vehicle \cite{dey2024multi}. It follows a dimming-and-glowing pattern at 0.8 Hz to communicate yielding, and solid light to indicate the intention to keep driving.

\subsection{Feature Activation}
These designs convey awareness and intent through features activated at the following specific moments (see \fig~\ref{fig:figure3} and \fig~\ref{fig:figure6}):

\begin{itemize}
\item \textbf{Point A (16 m) – Pedestrian Awareness:} \texttt{popular-design} shows a pedestrian icon; \texttt{innovative-design} projects green zone rotating toward the pedestrian; and \texttt{baseline} activates ``eyes,'' rotating toward pedestrian.
\item \textbf{Point B (5 m) – Yielding Intention:} the light bands of the \texttt{popular-design} change from cyan to green; the light band of the \texttt{innovative-design} turns green; and the \texttt{baseline} begins pulsing its middle light band.
\item \textbf{Point C – Crossing Facilitation:} the \texttt{popular-design} projects a zebra crossing; the \texttt{innovative-design} projects an arrow predicting walking direction; and the \texttt{baseline} remains unchanged from previous state.
\end{itemize}

\section{Results: Final Concept Evaluation}
\label{sec:Results_FinalConceptEvaluation}

A total of 131 participants took part of the final evaluation, all residents of European countries. Out of these, 13 entries were removed because of two or more missing key presses, signaling a lack of attention. The final cohort had an average age of 38.2 years (SD: 12.81), with a gender distribution of 1 non-binary, 59 female, and 58 male participants.
Out of these, 58 used a private vehicle for transportation, 37 Public transportation and 23 Walking/Cycling. Regarding autonomous vehicle acceptance, the general feeling was slightly above neutral (M: 3.15; SD: 0.98, in a 1-5 scale). 
The median study duration was 10 minutes, and participants were compensated at a rate of \pounds $8.67$ per hour for their involvement.

\subsection{\textbf{Response Time}}
We conducted a repeated-measures ANOVA to compare mean reaction times across the three eHMI conditions. The assumption of normal distribution was assessed with Shapiro–Wilk tests, which showed that response times for the \texttt{innovative-design} deviated from normality, while the other conditions met the assumption. A Levene test confirmed homoscedasticity. Given the large sample size, we proceeded with the repeated-measures ANOVA, despite one condition's departure from normality. The analysis revealed a significant main effect of eHMI condition, $F(2, 216) = 3.80, p = .024, \eta^{2} = .009$, with Greenhouse-Geisser corrected degrees of freedom ($\varepsilon = 0.975$). Post-hoc Bonferroni-corrected t-tests showed that \textit{response times were significantly faster for the} \texttt{popular-design} than for the \texttt{baseline} condition, $t(108) = -2.77$, $p = .020$, $Hedges'g = -0.23, BF_{10} = 3.93$. No significant differences were found between the \texttt{popular-design} and \texttt{innovative -design}, or between the \texttt{innovative-design} and \texttt{baseline}. These results are illustrates in left side of \fig~\ref{fig:figure7}. 

The center of \fig~\ref{fig:figure7} also shows response times across the full video sequence by plotting the percentage of participants who pressed the key. At point B (yielding intention), the \texttt{popular-design} shows a clear inflection, with more participants feeling safe, linked to the projection of green “safe zone,” a feature frequently proposed during the crowdsourcing phase. At point C (crossing facilitation), the \texttt{innovative-design} produces a sharp increase, reflecting the use of an arrow projection of the pedestrian's path, which also echoes recurring symbol-based proposals from the ideation phase.

\begin{figure*}[t]
    \centering
    \includegraphics[width=0.99\linewidth]{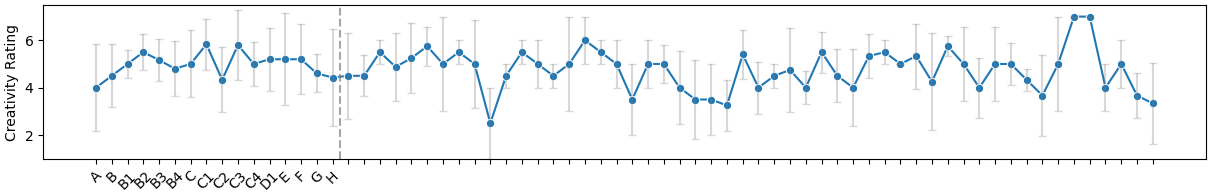}
    \caption{Participant creativity ratings shown in chronological order. Labels are removed for clarity, except for the design Concepts on the left of the plot, such as Concept A, Concept B, Concept B1, and others, which are described in \sect \ref{sec:Results_Collaboration_Feedback}.}
    \label{fig:figure8}
    \Description{The figure presents a line plot of creativity scores assigned by participants to various submitted ideas, showing a fluctuating, noisy distribution.}
\end{figure*}

\subsection{\textbf{User Experience}}
The User Experience Questionnaire scores violated assumptions of normality and homoscedasticity; therefore, we used a Friedman test as the non-parametric equivalent of a repeated-measures ANOVA. This analysis revealed a significant effect of eHMI condition, $\chi^{2}(2) = 103.39, p < .001$. Post-hoc pairwise comparisons using Wilcoxon signed-rank tests with Bonferroni correction indicated a significant difference between the \texttt{popular-design} and \texttt{innovative-design}, $W = 1512.0, p = .00037, Hedges'g = 0.26$; between \texttt{popular-design} and \texttt{baseline}, $W = 296.5$, $p < .001$, $Hedges'g = 1.23$; and between \texttt{innovative-design} and \texttt{baseline}, $W = 444.0$, $p < .001$, $Hedges'g = 0.97$. These results are illustrated in the right side of \fig \ref{fig:figure7}.

\subsection{\textbf{Qualitative eHMI interpretation}}
Participants' interpretations of the three eHMI designs revealed clear differences in perceived clarity and safety. The \texttt{popular- design} was most often described as clear and intuitive, with green pedestrian symbols, projected crosswalks, and bus-mounted lights effectively communicated when it was safe to cross. Many reported increased confidence due to the vehicle's apparent acknowledgment of their presence: ``...it was super clear that the bus was about to stop in advance... visually safe and super interesting'' (P19). Some participants, however, found the multiple signals slightly overwhelming when projected to the sides of the vehicle (P20).

The \texttt{innovative-design} was also well received, especially for its ground projections and arrows, which conveyed safety despite minor initial confusion (P70): ``I could clearly see a sort of green beam in front of the approaching bus and a sign appeared on the road in green which signaled me it is safe now to cross'' (P70).

In contrast, the \texttt{baseline} was frequently judged unclear, with participants relying on vehicle behavior rather than signals, leading to hesitation and lower confidence: ``The blue lights did not give me the feeling that it was safe for me to cross... I waited until I noticed that the vehicle broke'' (P117). 

\section{Discussion}
We begin this section by revisiting our research questions, then broaden the discussion to the role of crowdsourcing in involving everyday users in developing emerging technologies.

\subsection{How does a collaborative, iterative crowdsourcing approach shape participants’ collective vision of eHMI design? (RQ1)}

As seen in previous creative-focused crowdsourcing studies, our method was able to encourage participants to build on earlier submissions or propose entirely new ideas. Based on self-reported sources of inspiration, $57\%$ of participants (38/67) explicitly built on previous contributions, while $42\%$ relied primarily on the initial concepts. More granularly, sixteen participants ($24\%$) built directly on earlier contributions, twenty-two ($33\%$) drew on newly introduced concepts, thirty-two ($42\%$) focused on the original basic concepts, and nine described their ideas as entirely new\footnote{These results are not adjusted for exposure, meaning later contributions were visible to fewer participants, creating a skew toward early concepts and amplifying the apparent dominance of initial concepts.}. The distribution of this data is shown in Table~\ref{tab:BuiltOn} in Appendix~\ref{app:SourceParticipantContributions}.

Participants frequently referenced earlier ideas in their comments as well. Fourteen participants directly mentioned features introduced by others, such as \textit{Bluetooth}, \textit{pedestrian counting}, \textit{pavement lights}, and \textit{projected safe zones}. Among the bigger highlights, seven participants referred to directional lighting or projections after these had been introduced as \textit{Concept B2}, indicating a clear influence from prior contributions. The expert feedback also shaped later ideas. Participants suggested omnidirectional LEDs instead of highly personalized signals, shifted projected messages from text to symbols, and introduced privacy-oriented solutions such as ``walls of light.'' 
Importantly, this integration of feedback did not appear to diminish creativity. As shown in Figure~\ref{fig:figure8}, creativity ratings varied consistently across iterations. Nonetheless, the thematic analysis conducted across iterations revealed a sharp decline in the introduction of entirely new modalities after the second iteration. This trend likely reflects both participants' preference for familiar elements and the natural tendency of user-centered design processes to converge on practical solutions (as observed in the distribution of motivation themes in Figure \ref{fig:figure11b} in Appendix \ref{app:ThematicAnalysis}).

Overall, collaborative crowdsourcing led participants to build on previous ideas while still contributing original variations. Their collective vision leaned strongly toward familiar and intuitive features, especially the use of conventional traffic colors and signals. This emphasis aligns with their belief that eHMIs should primarily reassure pedestrians and clearly communicate vehicle awareness and intent. This preference was also evident in the evaluation phase, discussed in the next section.

\subsection{How do (visual elements) of user-generated eHMI designs compare to prior ones? (RQ2)}
\label{sec:RQ2HowDo}

When examining the results of the comparison study, we find consistent patterns across both \textit{response time} and \textit{user experience} (as also reflected in the qualitative eHMI interpretations submitted by participants). The \texttt{popular-design} incorporating the most frequently suggested visual elements achieved the strongest performance, followed by the \texttt{innovative-design}, showing support for \textbf{H1}. 
However, we found no significant difference in response times between the \texttt{innovative-design} and the \texttt{baseline}, providing no support for \textbf{H1.a}. In contrast, significant differences emerged in user experience ratings (also reflected in the qualitative feedback) supporting \textbf{H1.b}. Finally, we found no support for \textbf{H2}, as the \texttt{innovative-design} consistently ranked second across all metrics.

A closer look reveals that the key factor driving quicker response times was the pedestrian icon, which clearly indicated pedestrian detection and had been repeatedly suggested during the creative phase. In contrast, the \texttt{innovative-design's} tracking light alone was insufficient, as participants confidently understood the message only after the arrow was displayed. The \texttt{baseline} design, which relied on more abstract visual elements, performed worst. 

These findings follow reported trends that abstract cues, such as lights alone are not clear enough \cite{brill2023external}. We suggest that the use of simple symbols  or text could significantly enhance clarity. At the same time, we acknowledge that icons or texts alone are not scalable solutions. In practice, a more effective approach may be to extend the \texttt{innovative-design} by combining its tracking light with pedestrian icons or to adapt the \texttt{baseline-design's} ``eyes'' into multiple icons indicating pedestrian locations.

Overall, the consistent pattern across this evaluation study is the effectiveness of familiar or standardized visual elements. However, this should not overshadow the value of creative contributions from everyday participants. The \texttt{innovative-design}, although less effective than the \texttt{popular-design}, still performed second-best, and slight modifications---such as projecting icons throughout---could have further improved its performance. Interestingly, human-like features, though common in prior research \cite{jaguar2018, chang2017eyes}, were rarely suggested during the creative phase and performed poorly in evaluation alongside abstract elements. These results provide a strong indication to limit such features and instead favor more familiar ones. 
Standard elements have the clear advantage of reducing cognitive load due to participants' prior exposure. Whether human-like or abstract features could achieve similar familiarity over time remains an open question, but doing so would likely require unnecessary extra steps compared to adopting standardized signals. 

\subsection{Additional Discussion Points}

We deliberately constrained participation to residents of European countries. This choice helped minimize potential conflicts between public preferences, traffic laws, and human factors, since expectations and norms among participants were more likely to align. Our goal was to reach a sample broad enough to yield meaningful insights, without being so extensive as to mask contextual differences. As discussed earlier (\sect \ref{sec:RelatedWork}), the aim of designing holistic eHMIs for AVs is not to produce a universal solution, but to generate insights that are applicable within localized contexts. 

Nonetheless, our sample remains skewed in several ways. Most participants were based in the United Kingdom, most reported private cars as their primary mode of transportation, and---assuming general demographics of crowdsourcing platforms---participants are likely to be literate, well-sighted, and relatively tech-savvy. While screening options on crowdsourcing platforms can mitigate some of these biases, challenges remain in adapting the method for greater accessibility and inclusivity. Importantly, as argued in \sect \ref{sec:PositioningOurApproachInPD}, we do not view our approach as a replacement for classic participatory methods, which are better suited to involving vulnerable or less-privileged user groups. Instead, we position it as a complementary method that can scale user-centered design.

This scalability also highlights the potential for adaptation within industry. While companies have previously experimented with crowdsourcing idea generation \cite{bayus2013crowdsourcing}, such efforts often relied on superficial approaches, asking participants to submit ideas without seeking a deeper understanding of users. We addresses this missing dimension by grounding contributions in user perspectives and expectations. The objective is not to extract isolated ideas for implementation, but to generate insights that can meaningfully guide development and implementation of technology.

Finally, an open question concerns legal and intellectual property ownership. While our study was guided by scientific inquiry and user-centered design, it remains unclear how contributions should be handled if a participant's idea were directly developed and commercialized. Given the automotive industry’s high safety and quality standards, the likelihood that an idea would move directly to production is low; moreover, all concepts in our study received substantial expert feedback. Nonetheless, such scenarios call for fair and transparent negotiations among workers, platforms, requesters, and policymakers to ensure equitable treatment.

\begin{figure}[b]
    \centering
    \includegraphics[width=1\linewidth]{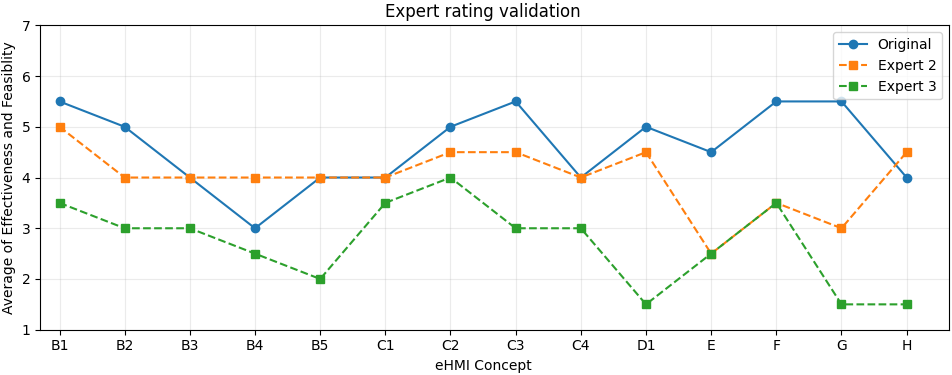}
    \caption{Comparison of the average Effectiveness and Feasibility ratings from the original expert and two additional experts.}
    \label{fig:figure9}
    \Description{The image displays a line chart titled Expert rating validation which compares the scores of three different evaluators across various design concepts.}
\end{figure}

\section{Limitations}
\subsection{Expert Feedback and Evaluation}
A key limitation of our work lies in the potential biases introduced through expert-led processes. For instance, the stopping criterion required judging whether an idea was novel or simply a ``substantial refinement'' versus a ``minor tweak''. This distinction is inherently subjective, and although we initially attempted to reduce bias using creativity ratings, this approach quickly proved ineffective. 
Similarly, expert evaluation of concepts may have introduced further bias, as judgments are inevitably shaped by personal perspectives and experiences.

To check for potential bias, we reviewed the expert's feedback. For effectiveness, high scores consistently highlighted that personalized communication and allocentric information can support safety and scalability, while low scores focused on ambiguity, unclear messaging, and misdirected communication. Overall, clarity received more weight than safety. Feasibility scores were also internally consistent, though slightly less than effectiveness scores. Low feasibility scores often cited GDPR, legal constraints, and scalability limits. High scores referred to concepts that are technically achievable but dependent on specific conditions.

Future implementations should involve at least two experts. In our case, limited resources forced us to consult two additional researchers outside our group to contextualize the original expert's ratings. The second expert is an assistant professor with more than ten years of experience in the field, and the third expert works as an industry leader at a European automotive company. Spearman correlations between the original expert and each additional expert showed weak positive relationships for the average of effectiveness and feasibility: with Expert 2 ($r = 0.11,; p = 0.723$) and with Expert 3 ($r = 0.23,; p = 0.441$). 

As shown in Figure \ref{fig:figure9}, the largest disagreement across all experts occurred for \textit{Flag Barriers} (Concept G). Design concepts where the new experts diverged from the original expert were Concept E (Robot) and Concept F (Bluetooth). Expert 3 also provided lower ratings overall. When applying the process described in Section \ref{sec:ResultsFinalDesigns-InnovativeDesign}, Expert 2's ratings would remove the \textit{directional light} (Concept B2) and add \textit{light pavements} (Concept C2) and \textit{eye-like features} (Concept H). Expert 3's ratings would remove \textit{future path projection} (Concept C3) and add \textit{projected icons or text} (Concept C1). Notably, all experts consistently rated the 360 band light (Concept B1) highly. 

Future work should revisit how feedback from multiple experts is aggregated and how it is integrated with user input. These differences, however, do not change the results and discussion of RQ2.


\subsection{Crowdsourcing Participants and Platform}

We also acknowledge limitations in our crowdsourcing platform. While we found no evidence of superficial responses or disengaged input---in fact, all but eight participants expressed interest in taking part in similar studies in the future---our sample was restricted to a single platform, Prolific. Prior work shows that Prolific workers tend to approach tasks more casually, treating them as a way to spend free time while earning some cash, whereas Mechanical Turk workers often rely on the platform as a primary income source \cite{oppenlaender2020creativity}. Testing our method across different platforms will therefore be an important step for future research. Nevertheless, our findings align with research showing that, although crowdworkers are rarely offered creative tasks, they are eager for more opportunities of this kind \cite{oppenlaender2020creativity}. This points to a promising opportunity to scale user-centered design and participatory methods through crowdsourcing.

Furthermore, we did not systematically track participants' design backgrounds. Only one participant volunteered relevant experience (``... I have also worked in a factory where AVs were first developed, so count me in!''). As a result, we could not examine whether professional expertise influenced contribution patterns. Nevertheless, we emphasize that people affected by a technology are experts in their own experience \cite{schuler1993participatory}. As potential pedestrians interacting with AVs, all participants brought relevant lived experience to the design of communication interfaces.

\subsection{Evaluation Scenarios}

Finally, we note limitations in using only low and high perceived risk scenarios (and only the later in evaluation), as these capture only a fraction of the situations pedestrians may experience in real life. We also focused exclusively on pedestrians, even though perspectives from other road users, such as cyclists \cite{Al2024light}, are equally important. In addition, our collaborative process generated multimodal concepts, but practical constraints restricted our evaluation to visual implementations in video simulations. As a result, conclusions about design performance (RQ2) apply specifically to visual eHMI elements. Future work should broaden this research to assess whether the findings extend to other perspectives and real-world conditions.

\section{Conclusion}
In this work, we demonstrated that collaborative crowdsourcing can effectively engage everyday users in the design of emerging technologies, using eHMIs for autonomous vehicles as a case study. Participants generated both creative and practical ideas, iteratively building on prior submissions while responding to expert feedback. Designs evolved from simple, single-modality cues to sophisticated multimodal systems integrating lights, projections, and audio, with a consistent preference for familiar and standardized elements that enhanced clarity and safety. Evaluation results, limited to visual elements only, showed that the most frequently suggested features outperformed baseline designs in both interpretability and user experience, while innovative concepts highlighted the potential for further improvement. 
While our case study focused on eHMI design in European contexts, the methodological framework---iterative ideation with expert feedback---may transfer to other emerging technologies where user expectations are still forming and standardization is absent. 
Our findings underscore the value of integrating crowdsourced creativity with expert guidance, highlight considerations around scalability, familiarity, and point to crowdsourcing as a promising avenue for participatory, user-centered design in future technology development.

\begin{acks}
This research was supported by the Horizon Europe EIC project \href{https://symaware.eu}{SymAware} under the Grant Agreement No. 101070802. We would like to thank the reviewers for providing valuable theoretical and practical critiques of the proposed method.
\end{acks}

\balance

\bibliographystyle{ACM-Reference-Format}
\bibliography{references}


\newpage
\appendix
\onecolumn

\section{Methodological Comparison}
\label{app:MethodologicalComparison}

\begin{table}[h]
\begin{tabular}{l|l|l|l}
\textbf{Dimension}                                         & \textbf{Traditional Survey}                                                            & \textit{\textbf{Our Approach}}                                                                              & \textbf{Traditional PD}                                                                           \\ \hline
Information Flow & \begin{tabular}[c]{@{}l@{}}Unidirectional\\ (participant → researcher)\end{tabular}    & \begin{tabular}[c]{@{}l@{}}Bidirectional (participants see\\ others' ideas + expert knowledge)\end{tabular} & \begin{tabular}[c]{@{}l@{}}Multidirectional (stakeholders,\\ designers, researchers)\end{tabular} \\ \hline
Iteration                                                  & Single response                                                                        & Building on prior ideas                                                                                     & Iterative co-design sessions                                                                      \\ \hline
Creativity Support                                         & \begin{tabular}[c]{@{}l@{}}Constrained by closed\\ or Likert-scaled items\end{tabular} & Open-ended with tree visualization                                                                          & Deep collaborative imagination                                                                    \\ \hline
Engagement                                                 & Minutes (shallow)                                                                      & $\sim$30 min. across activities (medium)                                                                    & Hours (deep, situated)                                                                            \\ \hline
Scale                                                      & 100 - 1000s participants                                                               & 100s participants                                                                                           & 5-25 participants typically                                                                      
    \end{tabular}
    \caption{Comparison of standard data collection methods with our proposed approach, as discussed in in \sect \ref{sec:PositioningOurApproachInPD}.}
    \label{tab:MethodologicalComparison}
\end{table}

\section{Source of Participant Contributions}
\label{app:SourceParticipantContributions}

\begin{table}[h]
    \centering
    \begin{tabular}{c|c|c}
   \textbf{Source} & \textbf{Type} & \textbf{Count} \\ \hline 
   A & Initial Concept & 9 participants built on this \\
   D & Initial Concept & 9 participants built on this  \\
   B & Initial Concept & 8 participants built on this  \\
   C & Initial Concept & 6 participants built on this  \\ \hline 
   C1 & Novel Concept  &  5 participants built on this  \\
   B1 & Novel Concept  &  5 participants built on this  \\
   B2 & Novel Concept  &  5 participants built on this \\
   B3 & Novel Concept  &  3 participants built on this \\
   C2 & Novel Concept  &  2 participants built on this \\
   F  & Novel Concept  &  2 participants built on this \\ \hline 
   P2 & Participant &  4 participants built on this \\
   P3 & Participant &  3 participants built on this \\
   P7 & Participant &  3 participants built on this \\
   P1 & Participant &  2 participants built on this \\
   P12 & Participant &  2 participants built on this \\
   P32 & Participant &  1 participants built on this \\
   P60 & Participant &  1 participants built on this \\ \hline 
   New Ideas & New Ideas &  9 participants built on this\\
    \end{tabular}
    \caption{Number of participants who built on different sources. \textbf{Note}: Results are not adjusted for exposure. Participants could only see earlier ideas, not all ideas. This means that ideas added later were visible to fewer participants.}
    \label{tab:BuiltOn}
\end{table}

\newpage
\section{Final Concept Visualization Tree}
\label{app:FinalConceptVisualizationTree}

\begin{figure*}[h]
  \centering
  \includegraphics[width=\linewidth]{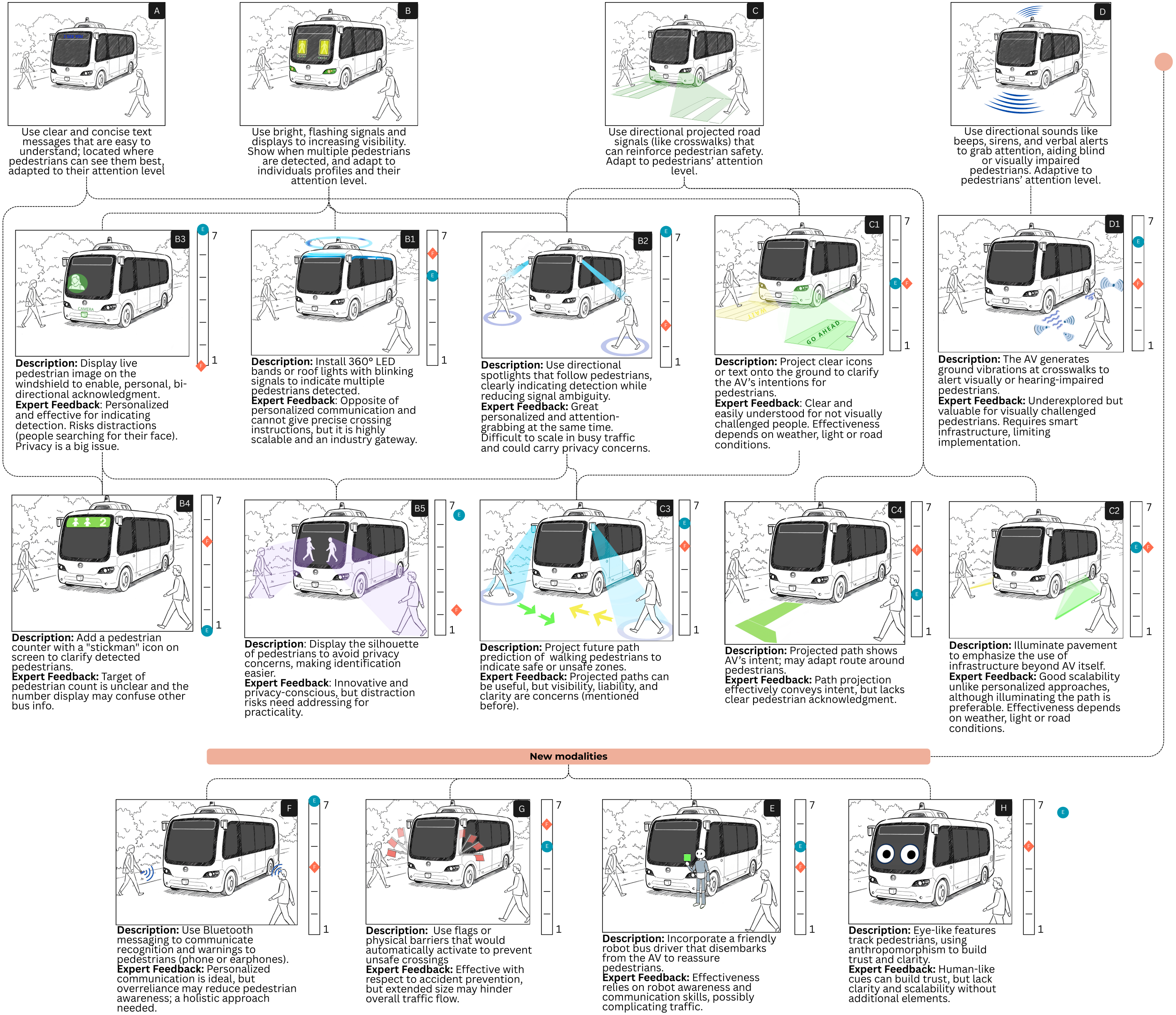}
  \caption{Final visualization tree presenting all concepts with brief descriptions and expert feedback, along with expert ratings for effectiveness ($\circ$) and feasibility ($\diamondsuit$) on a 1–7 scale, where 7 indicates the highest effectiveness and feasibility.}
  \Description{Description of Expert Analysis and Feedback ...}
  \label{fig:figure10}
\end{figure*}

\newpage
\section{Thematic Analysis - Code Distribution}
\label{app:ThematicAnalysis}

\begin{figure}[h]
     \centering
    \begin{subfigure}{0.5\textwidth}
         \centering
         \includegraphics[width=\textwidth]{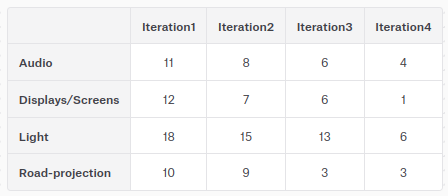}
         \caption{Design themes (counted once per participant).}
         \label{fig:figure11a}
        \Description{This figure is a table that presents the distribution of concepts across four development iterations, categorized by the type of sensory feedback they utilized. The table displays the frequency (or count) of ideas generated for each category in each successive iteration.}
    \end{subfigure}
    \begin{subfigure}{0.5\textwidth}
         \centering
         \includegraphics[width=\textwidth]{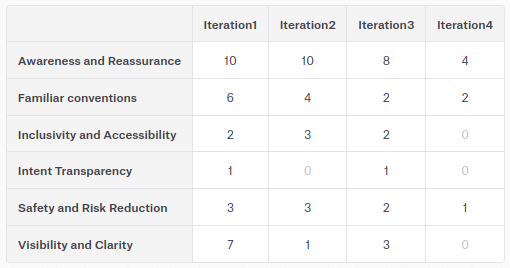}
         \caption{Motivation themes (counted once per participant).}
         \label{fig:figure11b}
        \Description{This figure is a table that presents the distribution of concepts across four development iterations, categorized by the type of sensory feedback they utilized. The table displays the frequency (or count) of ideas generated for each category in each successive iteration.}
     \end{subfigure}
     \begin{subfigure}[b]{0.95\textwidth}
         \centering
         \includegraphics[width=\textwidth]{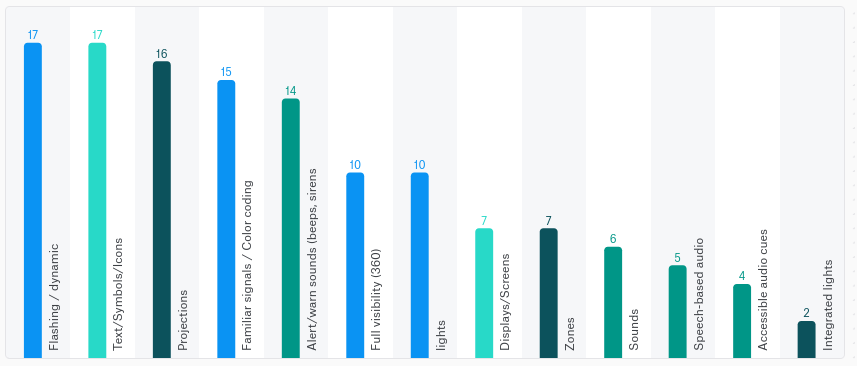}
         \caption{Distribution of individual thematic codes for design elements.}
         \label{fig:figure11c}
        \Description{The image displays a vertical bar chart that ranks various design elements and feedback mechanisms based on their frequency or count. The categories are arranged on the x-axis in descending order of frequency from left to right, with the specific count displayed at the top of each bar.}
     \end{subfigure}
    \caption{Distribution of individual codes and themes across all 4 creative iterations. Generated using the ATLAS.ti Web.}
    \label{fig:rankings}
\end{figure}

\end{document}